# Moisture Effects in Medium-Voltage Underground Distribution Systems: a perspective overview

Diana Enescu[1,2], Elena Otilia Virjoghe[4], Simone Vincenzo Suraci[5], Luigi Calcara[7], Radu Porumb[6], Lucia Rosso[2], Giulio Beltramino[2], Valentina Picco[3], Andrea Mazza[3*]

[1]Electronics, Telecommunications and Energy Department, Valahia University of Targoviste, Targoviste, Romania
[2] Physical Thermodynamics Unit, Istituto Nazionale di Ricerca Metrologica (INRiM), Torino, Italy
[3] Energy Department "Galileo Ferraris", Politecnico di Torino, Torino, Italy
[4] Automatics and Electrical Engineering Department, Valahia University of Targoviste, Targoviste, Romania
[5] Laboratory of Innovative Materials for Electrical Systems - Department of Electrical Engineering, University of Bologna, Bologna, Italy.
[6] Power Engineering Systems Department, National University of Science and Technology Politehnica Bucharest, Bucharest, Romania
[7] Department of Electrical and Energy Engineering, University of Roma "La Sapienza", Roma, Italy

[*]Corresponding author: andrea.mazza@polito.it

***Abstract* —** The distribution system is becoming a fundamental part for enabling the energy transition. The fault causes may be several and of different nature, embracing a variety of specializations. This review aims at providing an integrated overview of moisture effects across underground Medium Voltage (MV) cable systems and their accessories. The study collects both literature contributions and practical experiences derived from distribution system operators and research projects on the topic. The findings highlight that the choice of the insulation material is fundamental for mitigating the various consequences of the moisture ingress, linked to chemical and physical processes, as well as mechanical, thermal, and electrical factors, all of which are analysed in detail in the paper. The paper examines the effects of moisture on different types of materials, both traditional and innovative, and discusses diagnostic techniques to detect moisture ingress from the surrounding environment as well as internally generated moisture. The path towards increasing MV distribution reliability, reducing outages, and support the energy transition with more robust distribution infrastructure, is hence achievable through improving moisture management across design, installation, monitoring, and maintenance.

ACRONYMS

| | |
|---|---|
| AC | Alternating Current |
| ATH | Alumina TriHydrate |
| BN | Boron Nitride |
| CTD | Capacitance and Tan-Delta Testing |
| DC | Direct Current |
| DCP-XLPE | DiCumyl Peroxide-XLPE |
| DERs | Distributed Energy Resources |
| DFR | Dielectric Frequency Response |
| DSC | Differential Scanning Calorimetry |
| DSO | Distribution System Operator |
| EDS | Energy Dispersive X-ray Spectroscopy |
| EPDM | Ethylene-Propylene-Diene Monomer |
| EPR | Ethylene–Propylene Rubber |
| FDR | Frequency-Domain Reflectometry |
| FDS | Frequency-Domain Spectroscopy |
| FEP | Fluorinated ethylene propylene |
| FTIR | Fourier Transform Infrared Spectroscopy |
| HDPE | High-Density PolyEthylene |
| HEPR | Hard EPR |
| HMWPE | High–Molecular-Weight Polyethylene |
| HV | High Voltatge |
| IRT | Infrared Thermography |
| LDPE | Low-Density PolyEthylene |
| LV | Low Voltage |
| MV | Medium Voltage |
| PD | Partial Discharge |
| PDC | Polarisation-Depolarisation Current |
| PDD | Partial Discharge Detection |
| PE | PolyEthylene |
| PILC | Paper Insulated Lead sheath Cable |
| POS | Polyhedral Oligomeric Silesquioxanes |
| PRPD | Phase-Resolved PD |
| PSA | Pulse Sequence Analysis |
| PTFE | PolyTetraFluoroEthylene |
| PVC | PolyVinyl Chloride |
| PWA | Pulse Waveform Analysis |
| NIRS | Near-Infrared Spectroscopy |
| SEM | Scanning Electron Microscopy |
| SIED | Stress-Induced Electrochemical Degradation |
| SiR | Silicon Rubber |
| Si-XLPE | Silane-XLPE |
| TEM | Transmission Electron Microscopy |
| THz-TDS | Terahertz Time-Domain Spectroscopy |
| TI | Thermal Imaging |
| TMPTA | Trimethylolpropane Triacrylate |
| UV | Ultra Violet |
| WTR | Water-Tree-Retardant |
| XLPE | Cross-Linked PolyEthylene |

# 1 Introduction

The power system infrastructure is one of the most extensive and complicated infrastructure designed and built so far: it has a continental scale extension and includes a huge number of components and devices, spanning from Low Voltage (LV) (used in the last mile) to High Voltage (HV) ones (required for long-distances), with Medium Voltage (MV) networks enabling regional distribution and local electricity delivery [1].

In recent years, MV networks have been facing a paradigm change. Traditionally designed as passive infrastructures supplying MV loads and LV grids, MV networks are increasingly becoming active networks integrating distributed energy resources (DERs), including distributed generation, storage, and load-management systems [2]. In this context, reliable and resilient MV networks are essential for the implementation of energy-transition and decarbonisation policies. In many countries, MV networks are *underground* systems, hence composed of power cables (about 30 million km of MV cables estimated worldwide [3]) and MV joints.

Power cables must operate under different environmental and operational conditions and therefore require suitable electrical, mechanical, thermal, and chemical properties. Table 1 summarizes the principal cable-design requirements and their expected benefits [4].

***Table 1 Requirements for power cable design (adapted from [4]).***

| Property | Requirements | Benefits |
|---|---|---|
| *Flexibility* | Adaptable to underground installation, including narrow or curved spaces, remaining intact | Reduced installation time and reduced damage during installation |
| *Mechanical Strength* | Strong enough to resist pulling forces during installation | Prevention of breakage during pulling or high-tension stress |
| *Optimal Length* | Longer cable reduces the number of joints | Enhanced reliability and reduced maintenance costs |
| *Robustness against impacts during natural disasters* | Designed to endure shocks, vibrations, and impacts | Even in harsh conditions, the cable remains functional |

| Electrical characteristics | Low conductor resistance, strong insulation, minimal thermal and dielectric losses, and withstand overcurrents | Max efficiency and min energy losses |
|---|---|---|
| *Chemical characteristics* | Resistant to moisture, chemicals and harsh conditions; high thermal stability against ageing; fire-resistant with minimal smoke production in case of fire | Long-term durability in harsh and corrosive environments |

The expected service life of MV cables is commonly estimated to be approximately 30–40 years [5]. However, their actual service life may be reduced by severe operating conditions, including elevated temperatures, moisture ingress, and mechanical stresses occurring during installation and operation. These degradation processes may increase maintenance requirements and reduce the overall reliability of MV cable systems [6], [7].

Cable ageing is the gradual and generally irreversible deterioration of materials and functional properties resulting from prolonged exposure to environmental, mechanical, thermal, and electrical stresses [8]. The combined action of these stresses may accelerate insulation degradation and increase the probability of premature failure [9]. Among them, moisture is particularly important because it can affect insulation materials, metallic components, cable interfaces, joints, and substation equipment. Moisture can act as a *primary aging factor* by penetrating into insulation systems through damaged sheaths, interfaces, microcracks, or defects generated by mechanical, thermal, or environmental stresses. Once inside the insulation, moisture promotes the initiation and growth of defects such as voids and water trees, progressively weakening the dielectric properties of the material. Moisture also acts as a *secondary aging factor* by accelerating corrosion of metallic components, degradation of screens and sheaths, interfacial separation, and local mechanical deformation. These processes distort the electric-field distribution, increase localized electrical stress, and favor the occurrence of Partial Discharges[1] (PDs) and electrical treeing (due to technological imperfection and potentially appearing also without moisture [11]), which further weakens the insulation.

Moisture in the surrounding soil also affects cable thermal performance. Soil moisture content influences thermal resistivity and therefore the heat dissipation and ampacity of underground cables [12], [13]. Under unfavourable conditions, cable heating may cause moisture migration and the formation of dry zones with increased thermal resistivity [14], [15], [16]. This can increase cable operating temperature and further accelerate insulation ageing. The thermal behaviour of underground cables is addressed in established calculation methods such as IEC 60287-1-1 [17].

Although several studies have examined individual electrical, thermal, mechanical, or environmental stresses, a comprehensive treatment of moisture must consider their interactions across cables, joints, insulation materials, and substation equipment. This review therefore provides an integrated overview of moisture effects in MV distribution systems, combining material-level degradation mechanisms, diagnostic and ageing-assessment techniques, practical measurements, and experience reported by distribution system operators (DSOs).

The paper is organised as follows. Section 2 recalls the fundamentals about the components of the distribution systems, i.e., cables and cable joints, presenting the types and their evolution. Section 3 presents an overview on the types of insulation materials used for cables and joints and their sensitivity to the moisture. Section 4 shows a summary of the diagnostic techniques for cables and their ability to detect moisture permeation in insulation. Section 5 discusses some practical experiences, based both on an actual measurement setup and from information collected from DSOs. Finally, the last section contains the concluding remarks.

## 2 Overview Technological Aspects of Cables and Cable joints

### *2.1* Cables

Medium-voltage (MV) power cables typically operate in the 1–36 kV range. The main technologies are oil-impregnated paper-insulated lead-sheathed cables (PILCs) and extruded solid-dielectric cables. The latter, now widely used in urban distribution networks, employ polymeric insulation. Solid-dielectric MV cables were introduced in the early 1950s with butyl rubber and thermoplastic high-molecular-weight polyethylene (HMWPE). PILC remained dominant until the mid-1960s, when it was progressively replaced by HMWPE, cross-linked polyethylene (XLPE), and ethylene–propylene rubber (EPR) [18].

MV cables, as shown in Figure 1, have a multilayer structure: each layer provides a specific electrical, mechanical, or environmental protection function. The central stranded copper or aluminium conductor (1) carries the current. A semiconductive conductor screen (2) surrounds it, smoothing conductor irregularities and ensuring a more uniform electric-field distribution within the insulation. The main insulation (3), typically XLPE and EPR, is the most critical layer and separates the energised conductor from the grounded outer layers; its thickness depends on the rated voltage and service conditions. An outer semiconductive screen (4) provides a smooth interface between the insulation and metallic screen. The metallic screen (5), typically made of copper wires, copper tape, or a lead sheath, carries capacitive and fault currents, maintains the insulation surface at ground potential, and provides electromagnetic shielding. Finally, an outer sheath (6) made of polyethylene (PE), polyvinyl chloride (PVC), or another polymer protects the cable against moisture ingress, chemicals, abrasion, and mechanical damage.

[1] A PD, according to IEC 60270 [10], is "*the localised electrical discharge that only partially bridges the insulation between conductors and which can or cannot occur adjacent to a conductor*". PDs arise due to local electrical stress concentrations.

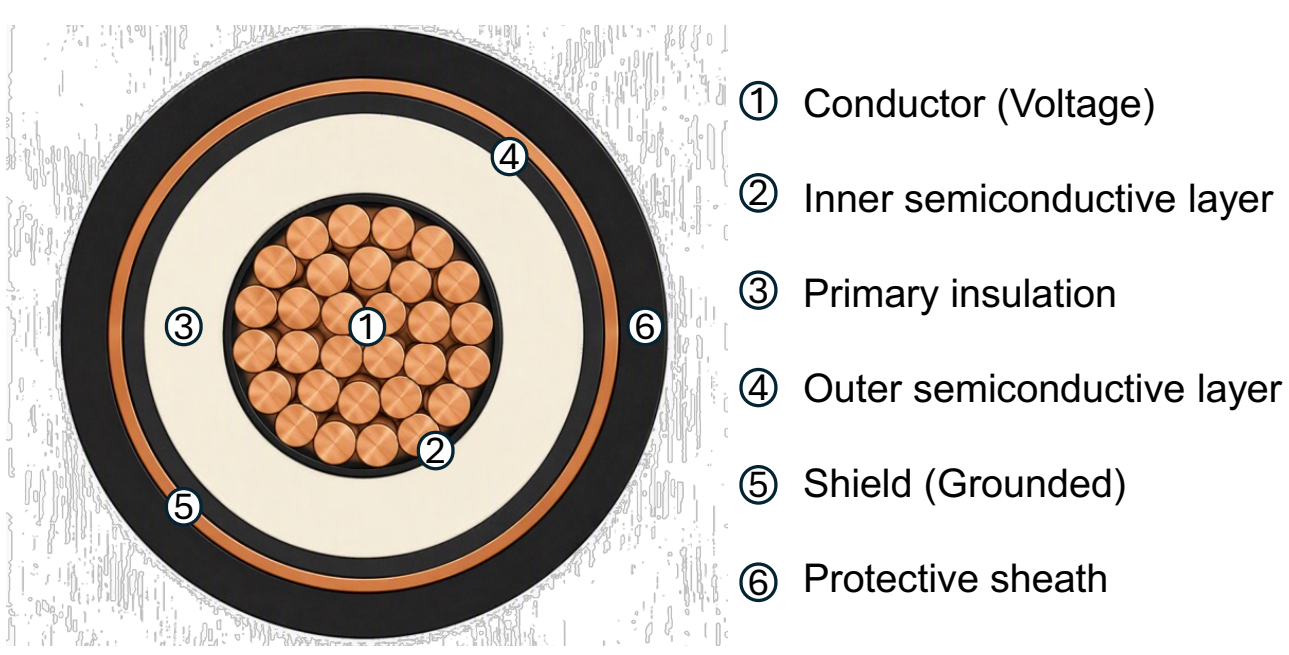

***Figure 1 Sketch of a MV underground distribution cable: main components.***

### *2.2* CABLE JOINTS

MV cable lengths are limited by reel capacity and are typically a few hundred metres, generally up to about 300 m depending on conductor cross-section. Feeders exceeding this length therefore require cable joints. Joints are often identified as a major source of cable-system failures [19][20], because their reliability is generally lower than that of the cable itself. Their multilayer structure combines insulating materials with different electrical properties, such as XLPE and ethylene-propylene-diene monomer (EPDM), producing non-uniform electric-field distributions and local field enhancement at material interfaces.

MV joints are normally assembled on site to connect adjacent cable sections. Even when premoulded slip-over designs are used, installation still involves manual operations. Imperfections at the interface between the cable insulation and the joint insulation, including voids and air gaps, may initiate PD activity and progressively degrade the insulation [21]. Joint replacement also shortens the cable length and increases the number of joints along a feeder, thereby raising the overall probability of failure.

Joint technology depends on voltage level, insulation material, installation method, and installation period. Older resin and oil-impregnated-paper joints remain in service. Before the 1960s, when underground distribution was dominated by PILC (few rubber or plastic-insulated cables were designed for voltages above 10 kV), joints and terminations for early solid-dielectric cables were generally hand-wrapped using insulating tapes. Self-bonding or self-amalgamating polyethylene (PE) tapes, and in some cases resin-impregnated open-weave tapes, were used to limit PD in voids between overlapping layers. These techniques were subsequently replaced by prefabricated modular systems, which reduce field labour and improve factory-controlled quality [18].

Modern MV systems mainly use heat-shrink and cold-shrink joints. Heat-shrink joints employ expanded cross-linked polyolefin components that recover onto the prepared cable ends under controlled heating, providing electrical stress control, moisture sealing, and mechanical protection. Cold-shrink joints use pre-expanded silicone or EPDM tubes supported by a removable plastic core. Once the core is extracted, the elastomer contracts and applies uniform radial pressure to the cable insulation and connector. This eliminates heating equipment and reduces installation time and errors associated with uneven heating.

Both technologies nevertheless depend on correct installation. Elastomer relaxation may gradually reduce contact pressure, while inadequate surface preparation, misalignment, or sealing defects can facilitate moisture ingress and increase the likelihood of PD activity [22].

The structural configuration of cables and joints determines the main moisture-ingress pathways, whereas the resulting degradation depends on the interaction between water and the constituent insulating materials. The following section therefore examines the relevant physical and chemical processes.

## 3 TYPES OF INSULATION MATERIALS AND THEIR SENSITIVITY TO MOISTURE

### *3.1* DIELECTRIC *PROPERTIES OF WATER*

Water–polymer interactions can significantly affect the long-term dielectric performance and integrity of cable insulation. Water is a strongly polar molecule with a permanent dipole moment of 1.85 D. At room temperature, its DC conductivity is approximately $3\text{-}5\cdot10^{-5}$ S/m, compared with values close to $10^{-17}$ S/m for insulating polymers, while its relative permittivity real part ranges from approximately 1.7 to 81 depending mainly on frequency.

At frequencies higher than $10^{10}$ Hz, electronic and atomic polarization dominate the dielectric response, molecular polarization dominates the dielectric response, independent of (*a*) the dipolar momentum and (*b*) temperature. At frequencies lower than $10^{10}$ Hz, thus including 50 and 60 Hz, the dipolar polarization becomes predominant. The real part of water permittivity is approximately 81 at room temperature and decreases to about 55 at 100°C. Consequently, even small amounts of moisture in cable insulation may increase dielectric polarization and losses, modify charge transport, and locally distort the electric field.

From a molecular perspective, water in cable polymers is commonly classified into three forms:

1. *Free water*, located in voids and defects with limited interaction with polymer chains.
2. *Bound water*, hydrogen-bonded to polar functional groups, oxidation products, fillers, or additives.
3. *Clustered water*, formed by water molecules hydrogen-bonded to each other within microcavities or free volume.

Bound and clustered water can modify the polarization mechanisms of the polymer matrix. Under AC fields, dipole orientation and relaxation increase dielectric losses and may alter the local electric-field distribution. Under DC conditions, moisture can affect conductivity gradients, space-charge accumulation, and charge trapping, thereby influencing the overall field distribution and insulation reliability [23]-[27].

Over extended service periods, thermal cycling and electrical stress may enhance moisture migration and contribute to water treeing, electrical treeing, and insulation embrittlement, described in the following.

### *3.2 Insulation Failure Mechanism in Humid Environments*

#### 3.2.1 *Cause-Effect Chain*

A tentative cause–effect chain leading from water permeation to insulation failure is shown in Figure 2.

The process begins with (i) water diffusion into the insulation (due to prolonged exposure to a humid environment) and (ii) its accumulation in free-volume cavities, voids, and interfacial regions. Because water has higher permittivity and conductivity than the surrounding polymer, local moisture accumulation may cause (iii) dielectric-property variations and hence (iv) electric-field distortion at defects and interfaces, creating localized field enhancements. Under sustained electrical stress, (v) microchannels may form and facilitate moisture migration and ionic transport, creating favourable conditions for (vi) water-tree initiation and propagation (see Section 3.2.2). Water-tree degradation progressively reduces dielectric strength and may locally enhance the electric field. Under specific conditions, this can facilitate (vii) partial-discharge activity and electrical-tree inception. In severe cases, long-term degradation may lead to (viii) dielectric breakdown. However, water trees do not inevitably evolve into electrical trees. This transition depends on electric-field intensity, insulation morphology, moisture content, ageing history, defects, and contaminants. Water treeing should therefore be considered as a *degradation mechanism* that may facilitate electrical tree inception rather than an unavoidable precursor to it.

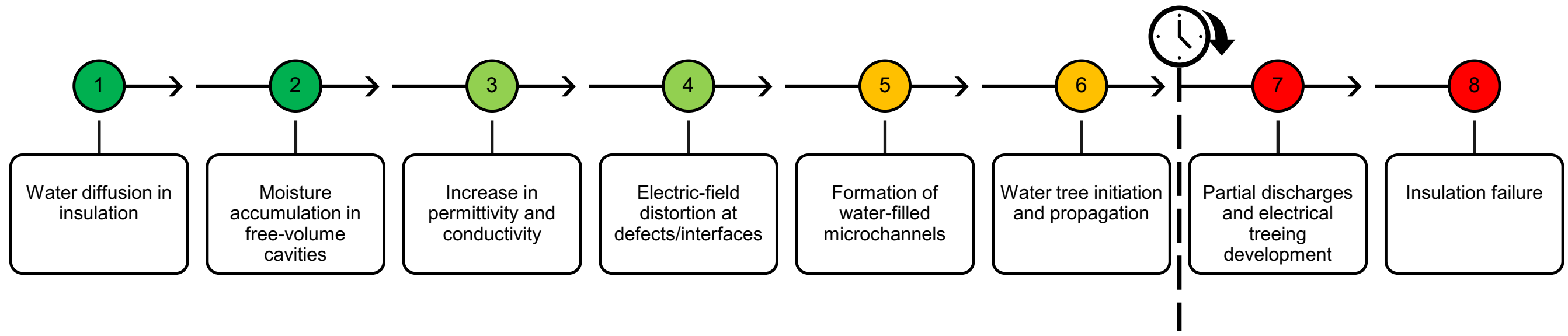

***Figure 2 Proposed cause-effect chain leading to insulation failure after water diffusion. The dashed line denotes a potential transition from water-tree formation to electrical treeing after long-term application.***

#### 3.2.2 *Water Treeing Formation and Propagation*

Water treeing is a slow electrochemical ageing process resulting from the combined action of moisture, electrical stress, and material imperfections. Water trees are microscopic branched degraded regions containing moisture-associated microcavities or channels, generally developing within the amorphous regions of polymeric insulation. The phenomenon has been widely studied in XLPE [28][29] but can also occur in EPDM, where suitable fillers may be used as tree-retardant additives [30][31]. Water tree initiation is commonly associated with contaminants, voids, protrusions, defects, or interfaces locally subjected to enhanced electric fields.

As shown in [32], water initially enters the polymer through sorption and diffusion processes governed by Henry's and Fick's laws. Especially during thermal cycling, it may become locally supersaturated and condense within microvoids or defects. Two principal theories have been proposed to explain water-tree initiation:

1. The *mechanical damage theory* [33][34]. Electrical and mechanical stresses generate or enlarge microcracks in the polymer. Under prolonged electrical stress, water diffuses and accumulates in these regions, promoting local mechanical and chemical degradation. Because of the permittivity difference between water and the polymer, Maxwell stresses may enhance crack growth and progressively interconnect microvoids into tree-like degraded structures that can extend over several millimeters within the insulation, incepting the treeing.
2. The *Stress-Induced Electrochemical Degradation (SIED)* [35][36][37]. Electrochemical reactions at the conductor–semiconductor interface, particularly in cables with aluminium conductors, generate corrosion products and hydrogen. The resulting porosity and microfractures in the conductor screen may initiate vented water trees. Their growth is subsequently influenced by electric-field strength, temperature, mechanical strain, ionic contaminants, and polymer morphology.

The combined action of these mechanisms progressively creates water-filled microchannels that reduce the dielectric strength of the insulation and may ultimately lead to electrical breakdown. Unlike electrical trees, water trees develop gradually over years or decades and are generally not associated with significant discharge activity during their growth. Nevertheless, they progressively alter the dielectric properties of the insulation by increasing moisture content, dielectric losses, and local conductivity, while reducing dielectric strength.

Two main morphologies are identified (Figure 3). Vented water trees originate at the conductor or insulation screen and grow into the insulation. Bow-tie water trees originate within the insulation bulk, generally around impurities, voids, or defects, and grow in opposite directions. Vented trees generally propagate more rapidly and are considered more harmful, although both morphologies can significantly degrade the insulation, and eventually lead to electrical treeing and cable failure if left unchecked.

Water trees rarely cause immediate failure, but extensive degradation may create favourable conditions for electrical-tree inception and eventual breakdown. Water-tree susceptibility depends on the chemical structure, polarity, degree of crosslinking, free volume, additives, and morphology of the insulating material. The following sections compare moisture transport and retention in XLPE, low-density polyethylene (LDPE), high-density polyethylene (HDPE), EPR, and EPDM, and discuss moisture ingress in cable accessories and interfaces.

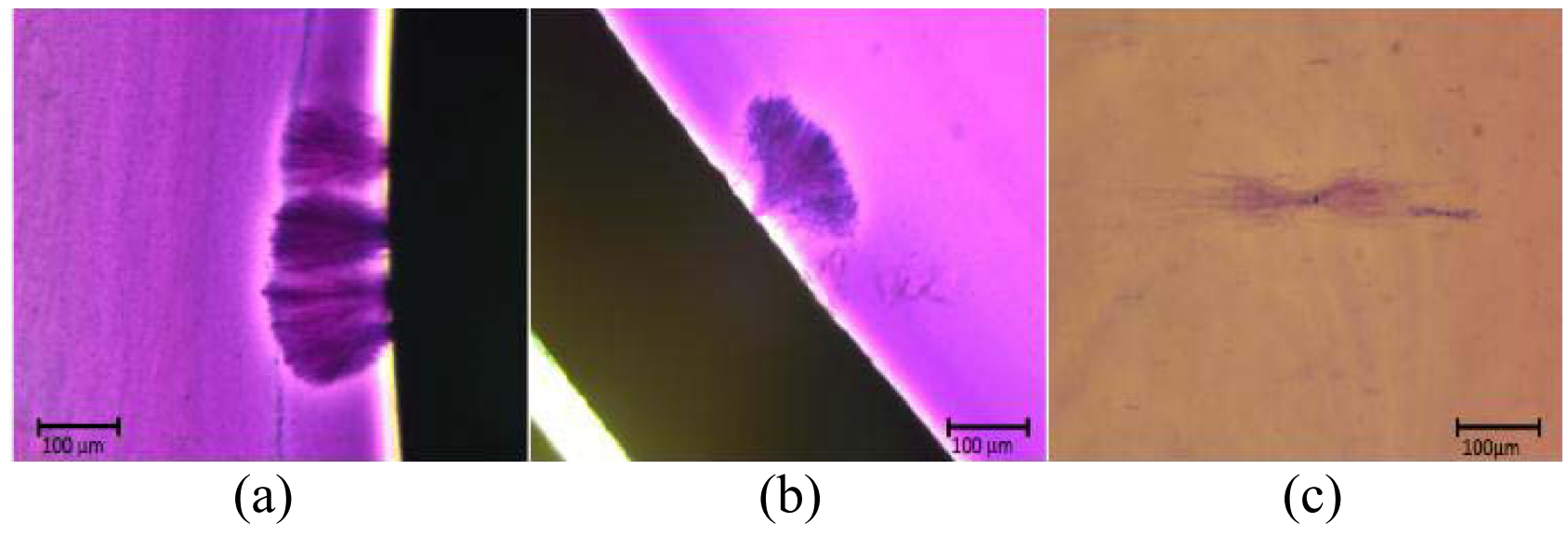

(a) (b) (c)

***Figure 3 Micrographs showing water treeings. (a, b) Vented trees, (c) Bow tie tree [38]***

### *3.3 Insulation systems*

#### 3.3.1 Crosslinked polyethylene

Polyethylene-based insulation is generally hydrophobic because its backbone consists mainly of non-polar carbon–carbon and carbon–hydrogen bonds which, as discussed in Section 3.1, do not provide the polar sites required for water bonding with the matrix. Nevertheless, water can penetrate through free volume, amorphous regions, defects, and crystalline–amorphous interfaces. Oxidation products generated during processing or ageing, including carbonyl, hydroxyl, and peroxide groups, introduce polar sites that increase water uptake [39]-[41].

As an example, Figure 5 compares the moisture content of neat silane-cross-linked polyethylene (Si-XLPE), Si-XLPE containing antioxidants, and Si-XLPE containing both antioxidants and a flame retardant. Antioxidants are commonly used in MV insulation compounds, whereas flame retardants are more typical of LV formulations. The samples were vacuum-dried for four days and then conditioned for one week at 20°C and 90% relative humidity. Moisture content was measured by Karl Fischer coulometric titration, while dielectric spectroscopy was used to assess the dielectric response before and after conditioning.

Unlike dicumyl peroxide-XLPE (DCP-XLPE), Si-XLPE contains silanol (Si–OH) groups that provide polar adsorption sites for water. Accordingly, even neat Si-XLPE showed increased moisture content after humidification. Phenolic antioxidants introduced additional polar groups and produced slightly higher residual moisture after drying. The largest increase in water uptake (+80%) was observed in the formulation containing the flame retardant alumina trihydrate ($Al(OH)_3$, ATH), whose hydroxyl-rich surface provides additional adsorption and retention sites.

Dielectric spectroscopy measurements were performed on these specimens. Figure 4 reports the real and imaginary parts of the relative permittivity at 0.1 Hz, 50 Hz, and 1 kHz. Si-XLPE exhibited a real permittivity of approximately 2.5, compared with about 2.3 for typical DCP-XLPE, consistent with the greater polarity introduced by silanol groups [42]. Antioxidants and ATH further increased the complex permittivity; the ATH-filled compound reached a real permittivity close to 4.5 at low frequency, also because ATH represented approximately 35 wt.% of the formulation.

Following humidification, an increase in the complex permittivity is observed for all materials, confirming the contribution of absorbed water to the overall polarization processes. The extent of this increase depends on the composition of the polymeric compound. Interestingly, the ATH-filled material exhibits only a moderate increase in permittivity despite showing the largest increase in water content (Figure 5). A possible explanation is that its dielectric response is already dominated by the high polarity of the ATH filler. Hence, the additional contribution from absorbed water is relatively low with respect to the base permittivity and it is therefore partially hidden.

By contrast, neat Si-XLPE and antioxidant-containing Si-XLPE showed a larger relative change, as the contribution of absorbed water was more evident against their lower initial permittivity.

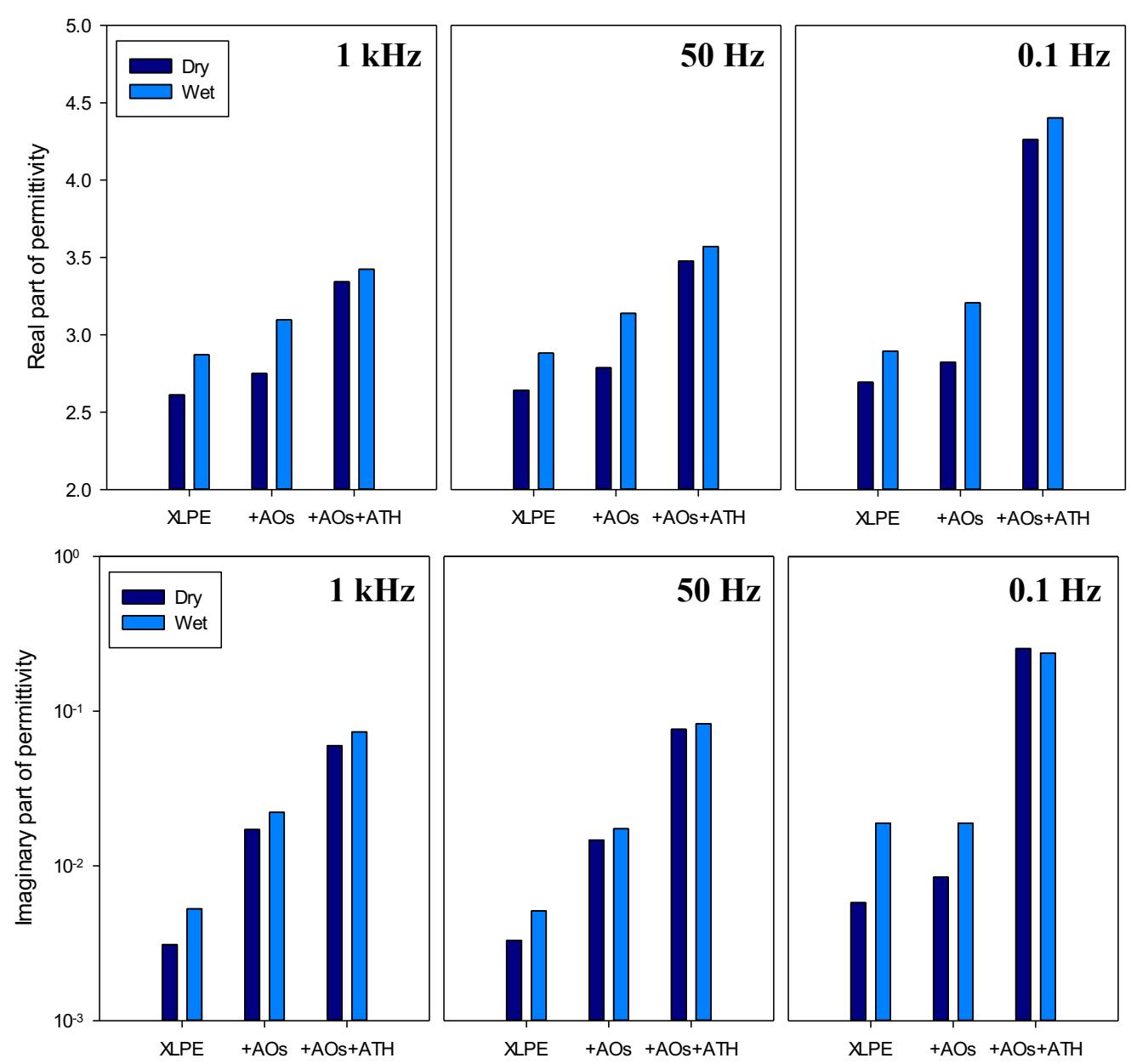

*Figure 4 Real and imaginary parts of permittivity for the three different XLPE-based compounds.*

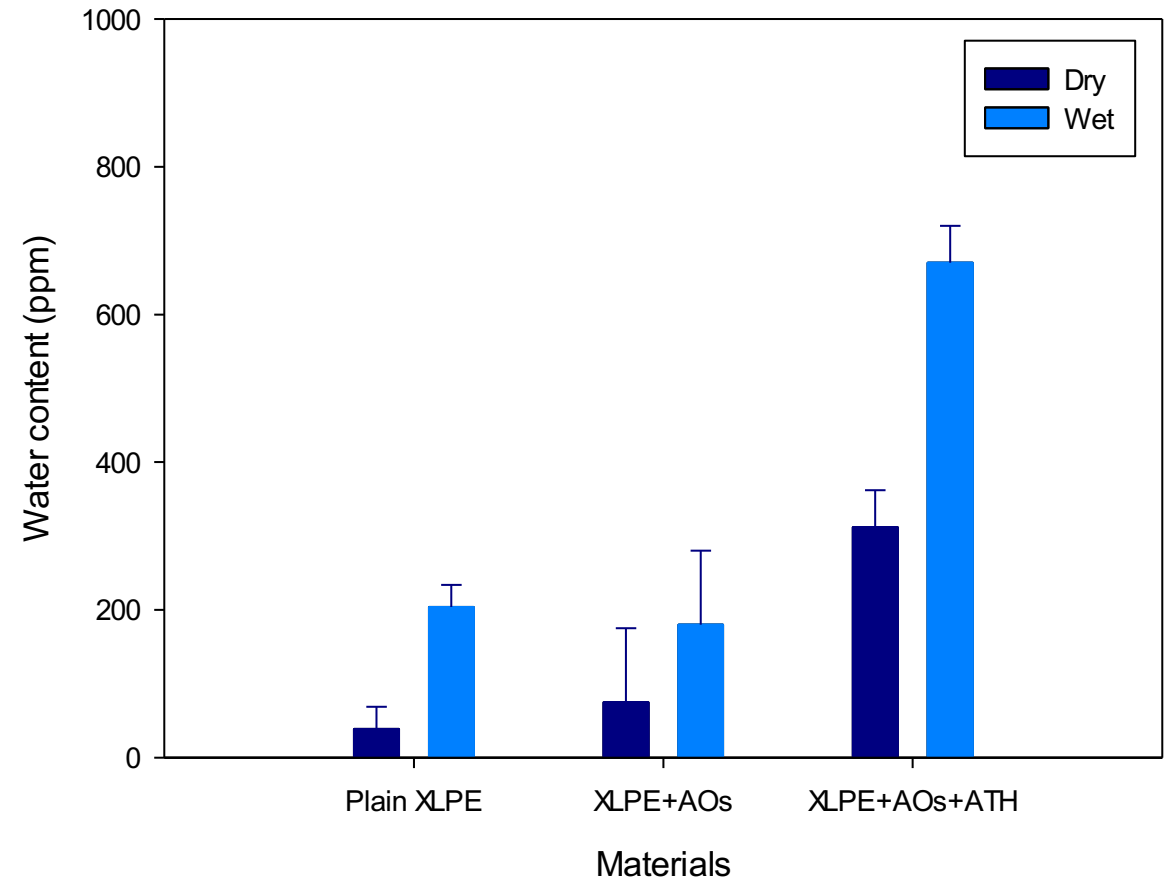

*Figure 5 Water content before and after climatic chamber treatment on XLPE-based polymers*

### 3.3.2 EPR AND EPDM RUBBERS

EPR and EPDM are used in underground cable insulation and accessories because of their good electrical, mechanical, and chemical properties, but they are not inherently flame retardant. EPR is commonly used in 6–36 kV MV cables and typically operates at temperatures up to 90 °C [8].

Water interaction is generally stronger in EPR and EPDM than in PE because these elastomers contain a less crystalline structure and often contain additives (e.g., cumyl alcohol from the peroxide byproducts), fillers (e.g., zinc oxide), and curing agents that introduce polar sites. Water molecules can associate with these polar groups through hydrogen bonding, leading to increased moisture uptake. Moreover, the large free-volume cavities of EPDM (190 $Å^3$, with radii 3.57 Å) can also accommodate more water (at least six water molecules) than PE and promote clustering. Reported equilibrium water uptake for EPDM immersed in distilled water at 70°C ranges from 1.2 to 2.7 wt.%, approximately 100–200 times the values reported for PE under comparable conditions [REF][43].

The EPDM specimens examined in [REF][44], after 720 h of water immersion, showed an increase in relative permittivity from approximately 2.4 to 3.4 and a decrease in volume resistivity from $6.3 \times 10^{14}$ to $1.0 \times 10^{14}$ Ω·cm, corresponding to an approximately 84% reduction (Figure 6). These results clearly indicate that water uptake significantly enhances the polarizability and electrical conductivity of EPDM, thereby altering its dielectric response and potentially affecting its long-term insulation performance.

A separate study monitored EPR-insulated cables (typically more protected against moisture due to the multi-layer structure) during two years of water immersion [REF][45]. As shown in Figure 7, insulation resistance remained relatively stable during the initial phase but decreased by more than two orders of magnitude after approximately 500 days, when moisture analysis indicated that the insulation was approaching saturation. The final resistance was of the order of $10^2$ MΩ, demonstrating that moisture accumulation can severely compromise the dielectric integrity of EPR insulation and significantly accelerate insulation aging.

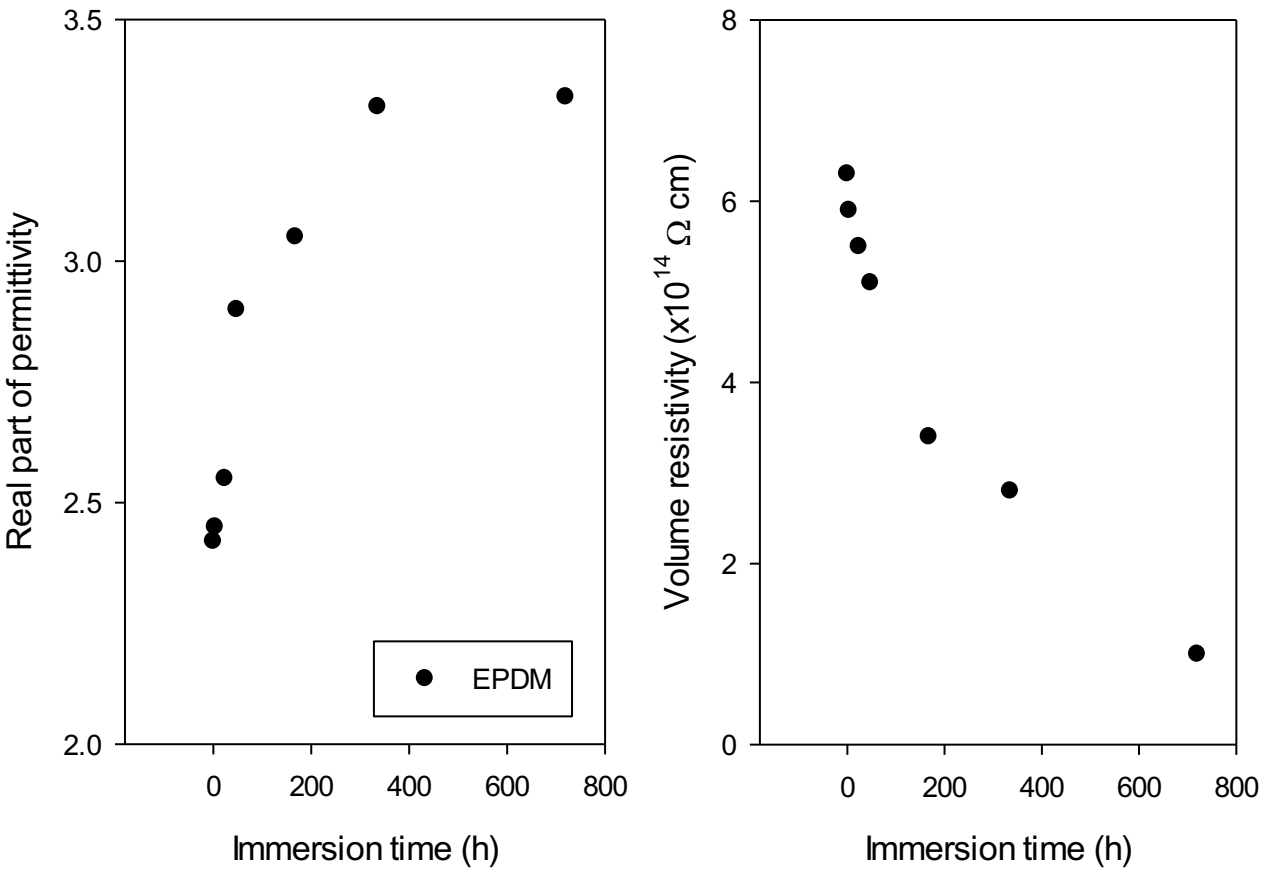

***Figure 6(a) Real part of permittivity and (b) insulation resistivity of EPDM sheets as a function of hours of immersion. Reproduced after [44]***

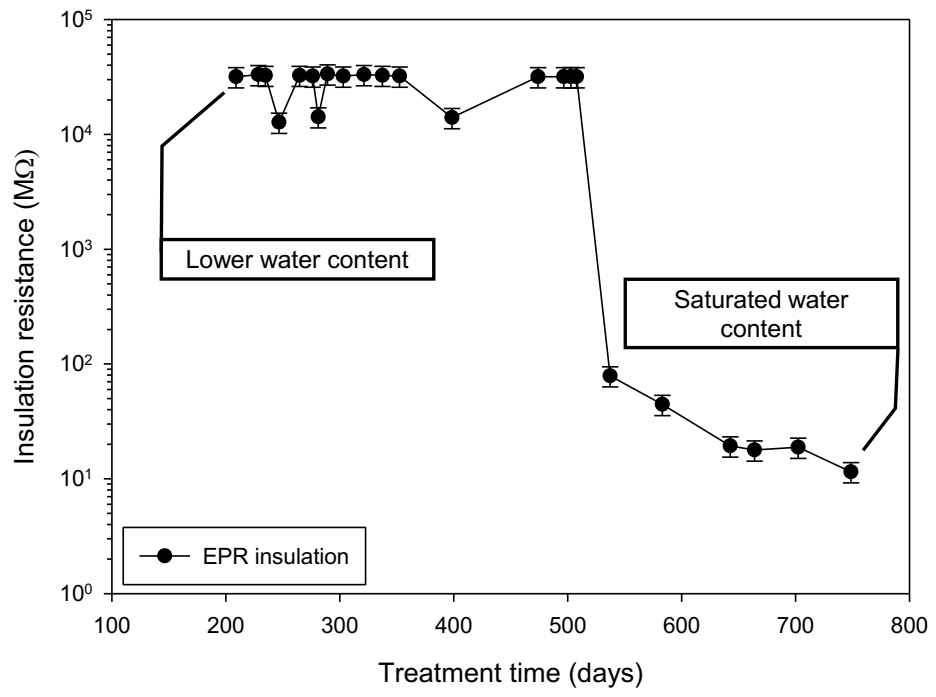

***Figure 7 Insulation resistance of EPR insulation as a function of days immersed in water. Reproduced after [45]***

Table 2 provides a comparative summary of the key properties governing water permeation in XLPE and EPR/EPDM insulation systems, together with their possible implications for cable performance and ageing.

*Table 2* **Summary of the key properties and effects of water permeation in XLPE and EPR/EPDM insulation systems.**

| | **XLPE** | **EPR/EPDM** |
|---|---|---|
| **Key characteristics** | - Semicrystalline, non polar backbone<br>- Very low moisture uptake<br>- Excellent dielectric properties<br>- Widely used in HV and MV distribution cables | - Amorphous elastomer<br>- Contains diene for vulcanization<br>- More polar sites and additives than XLPE<br>- Higher moisture uptake<br>- Used in cables and its accessories |
| **Water interaction** | - Free water – resides in voids, defects and amorphous regions<br>- Clustered water – water molecules bond with each other in microcavities<br>- Bound water – associates with polar sites (oxidation products, impurities or additives) | - Free water – resides in voids, defects and amorphous regions<br>- Clustered water – cavities are usually larger than XLPE, allowing higher water content.<br>- Bound water – associates with polar sites (within the polymer backbone) |
| **Water diffusion pathway** | - Diffuses mainly through amorphous regions<br>- Traps at defects, interfaces and contaminants (very low in the case of HV)<br>- Very low overall uptake (< 0.1 wt%) | - Easier diffusion due to bigger amorphous structure<br>- Water associates with polar groups and filler surfaces in the compound.<br>- Higher uptake than XLPE (~0.3 – 0.5 %wt). |
| **Effects** | - Increase in complex permittivity.<br>- Increase of DC conductivity.<br>- Increased PD activities.<br>- Moisture can initiate water trees. | |

### 3.4 *NEW MATERIALS AND ADVANCEMENTS*

Technological advances in insulation have led to the use of nanocomposites in power cables. Their relative permittivity, dielectric losses, PD behaviour, and resistance to water-tree formation depend largely on the type and concentration of nanofillers and additives dispersed in the polymer matrix, as well as on their surface treatment [46].

Nanocomposites used for underground cable insulation include materials based on XLPE, LDPE, PVC, elastomers such as EPDM and EPR, and fluoropolymers such as polytetrafluoroethylene (PTFE) and fluorinated ethylene propylene (FEP). Common nanofillers include $SiO_2$, $TiO_2$, $Al_2O_3$, MgO, and ZnO [47].

For example, on the one hand the introduction of silane-coated $Al_2O_3$ nanoparticles with a diameter of 50 nm was reported to reduce the electrical conductivity of LDPE nanocomposites by a factor of 50 compared with pure PE [48]. On the other hand, the addition of polyhedral oligomeric silsesquioxane (POS) fillers increased the thermal conductivity of LDPE by approximately 8% [49], while boron nitride (BN) particles increased it to approximately 1 $W \cdot m^{-1} \cdot K^{-1}$ at a filler concentration of 40 wt.% [50].

#### 3.4.1 *EFFECT OF NANOFILLERS IN WATER PERMEATION*

Nanofillers also influence the moisture behaviour of nanocomposite materials. The authors in [51] proposed a conceptual model describing the effect of water uptake on the DC conductivity of LDPE/MgO nanocomposites, as illustrated in Figure 8. Before nanoparticles are added (*a*), water molecules, ions, and other mobile polar species are assumed to be relatively uniformly distributed throughout the LDPE matrix, contributing to charge transport and electrical conductivity. At low nanofiller concentrations (*b*), nanoparticle surfaces provide energetically favourable adsorption sites. Water molecules, ions, and charge carriers progressively migrate towards these surfaces and become localized within the interfacial regions (*c*), reducing the concentration of mobile species in the surrounding polymer matrix. When the nanocomposite is exposed to a humid environment, additional water enters the material (*d*), increasing the DC conductivity. At higher nanofiller concentrations (*e*), nanoparticle aggregation decreases the effective surface area. Since adsorption occurs at the particle surface, fewer active sites are available to capture water and ions. Consequently, the beneficial purification effect becomes weaker, and the conductivity reduction is less pronounced. At sufficiently high concentrations (*f*), water-rich interfacial regions surrounding adjacent nanoparticles may overlap and create interconnected pathways for charge transport. The resulting percolation effect explains why an optimum filler concentration may exist rather than a continuous improvement with increasing nanofiller content.

Hydrophilic nanoparticles such as silicon dioxide ($SiO_2$) may promote moisture absorption. Studies [52] and [53] examined the dielectric behaviour of XLPE/$SiO_2$ nanocomposites in highly humid environments. Compared with pure XLPE, these nanocomposites exhibited higher moisture uptake because of the formation of water shells around the nanoparticles and changes in interparticle distances. Wet samples containing 5% $SiO_2$ also showed increased dielectric losses over the frequency range from 1 to $10^5$ Hz.

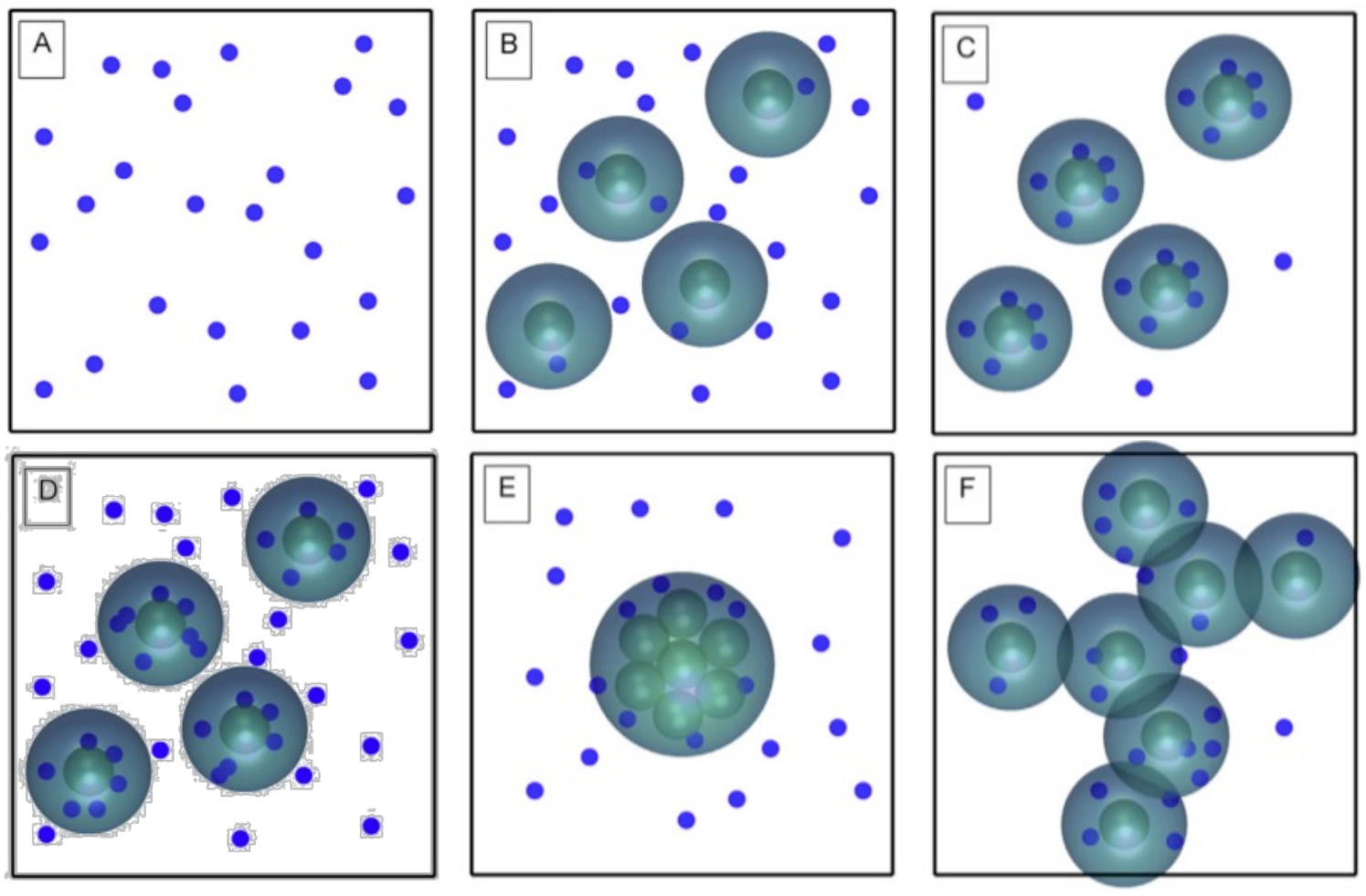

***Figure 8 Schematic description of the effect of moisture molecules on nanoparticles in nanodielectric materials LDPE/MgO [51].***

### 3.4.2 *IMPACT OF NANOFILLERS ON WATER TREEING*

Nanofillers can also modify water-tree inception and propagation. A mechanism proposed in [52] is based on the hydrophilic nature of silica nanoparticles. Similarly to conventional water-tree-retardant (WTR) additives, hydrophilic silica may modify the local distribution and transport of moisture within the insulation, thereby reducing the tendency for water-tree initiation and growth in XLPE/silica nanocomposites.

In [54], it has been experimentally investigated the impact of nano-$SiO_2$ particles on water tree growth in LDPE. The study showed that nano-$SiO_2$ could slow down the formation of water trees in PE and it was also found that hydrophilic nano-$SiO_2$ is more effective in reducing the growth of water trees than the hydrophobic variant. The influence of MgO nanofillers on the reduction of water trees in LDPE and XLPE was studied in [55]. It was observed that the development of water trees in nanocomposites slowed down, and this effect was amplified as the concentration of nano-MgO increased. Thus, it can be concluded that the addition of inorganic nanoparticles to PE leads to a slowdown in the formation of water trees.

In [56], the authors studied the effect of montmorillonite (Mnt) on the formation of water trees in XLPE. Mnt is a mineral clay composed mainly of hydrated aluminium silicate, with a layered structure that gives it special properties, such as water absorption and swelling capacity. The results showed a reduction in the length of water trees in the nanocomposite samples, with more pronounced decrease in materials with higher filler content. This improvement was attributed to the barrier effect of Mnt and its layered structure, which prevents the formation of water trees.

### *3.5 SUMMARY AND SOME PRACTICAL INDICATIONS*

Table 3 provideS a comparative overview of the dielectric, moisture-related, and water-treeing characteristics of polymer nanocomposite insulation systems. The collected results enable the identification of common trends regarding the influence of different nanoparticle fillers on the electrical and physical performance of cable insulation materials under humid operating conditions. The reported outcomes vary depending on nanoparticle type, concentration, surface treatment, and testing conditions. However, consistent observations can be drawn from the available experimental evidence, nevertheless highlighting the complex interactions between the polymer matrix, nanoparticles, and absorbed moisture, as well as their combined effect on dielectric behavior, water-tree resistance, and long-term insulation reliability

1. Nanodielectrics are normally characterized by values of permittivity different from the matrix, depending on the nanoparticle concentration and possible particle agglomeration. In some cases, increases in the permittivity of the nanocomposites were found due to the higher permittivity of the nanoparticles compared to the base polymer and the overlap of the interaction zones. Also, the permittivity of the nanocomposites depends on frequency and temperature [57].

i. $SiO_2$ and boron nitride (BN) nanocomposites exhibit better resistance against water tree growth by acting as a barrier that causes the branching and spreading of water tree channels. This process leads to better insulator behavior in high humidity environments [58]. Moreover, due to the hydrophobic capacity of BN nanocomposites, they are more suitable than those made of $SiO_2$ for use in humid environments, especially due to the lack of formation of a water film around the particles and due to the maintenance of dielectric properties ($\varepsilon_r$, tan $\delta$) in humid environments [59].

ii. $SiO_2$/XLPE nanocomposites are enhanced with the crosslinking agent TMPTA (Trimethylolpropane Triacrylate) which is used to develop TMPTA-s-$SiO_2$/XLPE nanocomposites with increases in mechanical strength, electrical strength and water tree resistance of $SiO_2$/XLPE nanocomposites. The reduction in thermal diffusion of water molecules is achieved by a higher degree of crosslinking which will form a denser network of molecular chains between the polyethylene lamellae, thus inhibiting the growth of water tree [60].

iii. Moisture absorption can increase the relative permittivity of nanocomposite insulation materials due to the presence of water molecules in the insulation. However, nanoparticles with modified surface (i.e., nanoparticles whose surfaces are treated with chemical substances to reduce water absorption) can reduce the effect of moisture on dielectric properties ($\varepsilon_r, \tan\delta$). This behaviour was observed in nanocomposite insulation materials containing titanium dioxide ($TiO_2$) and magnesium oxide (MgO) nanoparticles [61].

***Table 3 Effect of nanoparticle content on characteristics of different nanocomposites.***

| **Insulation + Nanoparticle (0.5÷5 wt%)** | **Relative permittivity, $\varepsilon_r$** | **Dielectric loss angle, tan $\delta$** | **References** |
|---|---|---|---|
| XLPE+$SiO_2$ | 2.3 ÷ 2.4 | $1 \div 3 \cdot 10^{-4}$ | [57], [58], [59], [60] |
| XLPE+$Al_2O_3$ | 2.3 ÷ 2.6 | $1 \div 4 \cdot 10^{-4}$ | [62] |
| XLPE+MgO | 2.3 ÷ 2.5 | $1 \div 4 \cdot 10^{-4}$ | [61], [63] |
| XLPE+BN | 2.3 ÷ 2.5 | $1 \div 3 \cdot 10^{-4}$ | [58], [59] |
| EPR+ $SiO_2$ | 2.8 ÷ 3.1 | $3 \div 7 \cdot 10^{-4}$ | [58], [59] |
| EPR+ $Al_2O_3$ | 2.9 ÷ 3.2 | $3 \div 8 \cdot 10^{-4}$ | [61] |

## 4 DIAGNOSTIC TECHNIQUES RELATED TO MOISTURE IN THE MV UNDERGROUND POWER CABLES

Understanding the physical effects of moisture on insulation systems enables the selection of appropriate diagnostic techniques and the correct interpretation of their outputs. A recent and complete review of the diagnostic techniques in MV underground power cables may be found in [39].

Moisture may either penetrate the cable system from the external environment (through defects in the sheath, joints, or terminations), or develop within the insulation as a consequence of ageing processes, such as water-tree growth. Table 4 summarizes the main diagnostic methods found in the literature and indicates their sensitivity to moisture ingress, and their typical field of application.

Microscopy laboratory characterization methods are reported separately in Table 5, because they generally require sampling of the cable material and are not suitable for on-site diagnostics.

***Table 4 Diagnostic techniques for moisture assessment in MV underground power cables.***

| **Category** | **Diagnostic Method** | **Description** | **Moisture permeation in insulation** | **Typical Application** | **References** |
|---|---|---|---|---|---|
| **Electrical and dielectric response techniques** | Insulation Resistance (IR) | Evaluates insulation resistivity, which decreases in the presence of moisture contamination. | ✓ | Condition assessment and moisture screening | [64] |
| | Capacitance and tanδ Testing | Monitors changes in capacitance and dielectric losses caused by moisture absorption and insulation ageing. | ✓ | Detection of moisture ingress and overall insulation degradation | [65] |
| | DFR/FDS (Dielectric Frequency Response / Frequency Domain Spectroscopy) | Analyses dielectric response over a broad frequency range to evaluate moisture content, dielectric ageing, and water-tree development. | ✓ | Global assessment of insulation condition | [66], [67], [68] |
| | PDC (Polarization-Depolarization Current) | Measures polarization and depolarization currents, allowing identification of conductivity and polarization changes associated with moisture. | ○ | Detection of moisture-related ageing phenomena | [69]-[71] |
| **Reflectometry techniques** | TDR/FDR | Identifies impedance discontinuities and changes in dielectric properties caused by moisture penetration, particularly near joints and accessories. | ✓ | Localization of moisture ingress paths and weak sections | [72] |
| **Partial discharge based techniques** | PD Analysis (including PRPD, PSA, and PWA) | Detects and analyses discharge activity associated with moisture-induced defects, water trees, and electrical trees. PRPD, PSA, and PWA provide complementary signal interpretation approaches rather than independent diagnostic methods. | ✓ | Identification and characterization of insulation degradation zones | [73]-[80] |
| **Spectroscopic techniques** | FTIR | Detects chemical modifications associated with oxidation, hydrolysis, and moisture-related degradation mechanisms. | ✓ | Assessment of chemical degradation of insulation materials | [81] |
| | NIRS | Uses near-infrared absorption characteristics to identify changes in molecular structure | ○ | Laboratory analysis of moisture-induced ageing | [82] |

| | | associated with moisture and ageing. | | | |
|---|---|---|---|---|---|
| | THz-TDS | Evaluates attenuation and scattering of terahertz waves within insulation materials, enabling early detection of moisture-related defects. | ✓ | Early-stage detection of degradation and water-tree development | [83] |
| **Thermal techniques** | Infrared Thermography (IRT) | Detects temperature anomalies associated with increased dielectric losses, insulation deterioration, and moisture-related defects. | ✓ | Non-invasive condition monitoring and fault screening | [84], [85] |

Legend: ✓: technique that proved to be influenced by moisture permeation, ○: technique that could be related to the moisture permeation if in combination with other techniques.

***Table 5 Microscopy laboratory characterization techniques (not suitable for on-site diagnostics)***

| **Technique** | **Description** | **Main Use** | **References** |
|---|---|---|---|
| SEM | Provides high-resolution images of insulation morphology and water-tree structures. | Post-mortem analysis and research investigations | [72], [86] |
| EDS | Determines elemental composition and chemical changes associated with insulation degradation. | Complementary laboratory assessment, typically coupled with SEM | [87] |
| TEM | Characterizes nanoscale morphological changes in degraded insulation materials. Requires extensive sample preparation and is unsuitable for field diagnostics. | Advanced laboratory research | [86] |

# 5 PRACTICAL FIELD EXPERIENCES

## *5.1* CABLE JOINTS FAILURES INVESTIGATION: AN EXAMPLE OF EXPERIMENTAL SETUP

This section aims to provide information about a measurement setup able to collect both moisture and temperature within a DSO facility.

The experimental set-up, whose deployment on several MV underground feeders started in summer 2014 [88], enabled monitoring (through direct measurements of environmental variables (temperature and moisture), cable and joint temperatures and load current) the variations in cable temperature and in the surrounding soil, and correlates these quantities with the evolution of ambient conditions over the year. Thermal-profile analysis also makes it possible to highlight the influence of precipitation on cable temperature, as well as the effect of solar radiation at different soil depths.

The experimental arrangement was equipped with thermocouple temperature sensors installed at representative positions and depths (Figure 9): at the ground surface (T1), at 0.5 m depth (T2), on the cable outer sheath at approximately 1 m depth (T3) and on its joint (T6), at the same depth as the cable but at a distance of non-influence of about 3 m (T4) and at 0.8 m below the cable (T5).

To characterise the soil water content, moisture sensors were installed both at the surface and underground, while the current carried by the cable was measured by means of a current transformer installed on the line.

All sensors were connected to a central datalogger, which enabled synchronous acquisition of the measurements, local storage, and subsequent transfer of the data to external media for post-processing. Installation at real operating sites allowed the thermal and environmental behaviour of cables and joints to be analysed across the different seasons [89].

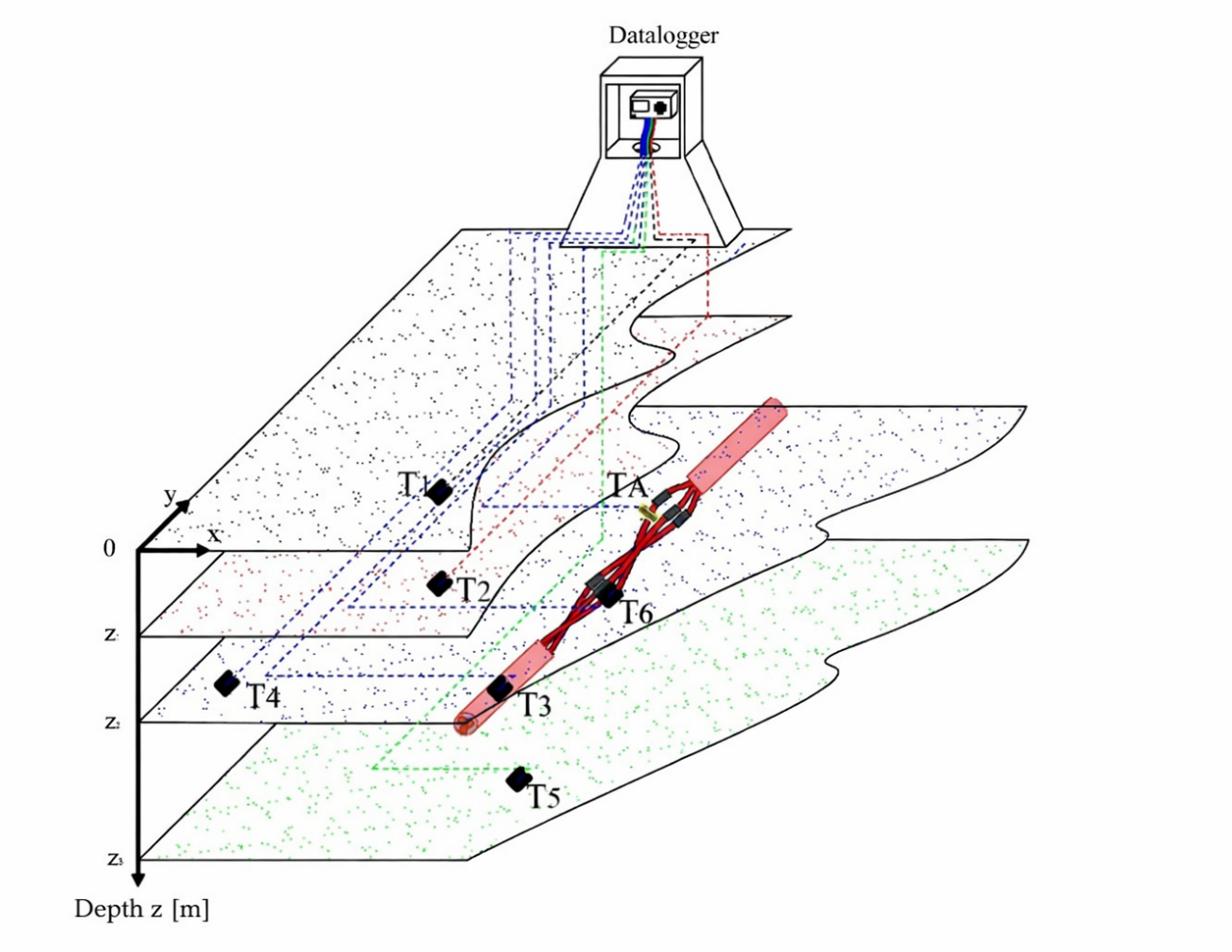

***Figure 9 Schematic representation of the underground experimental setup for thermal monitoring of an underground medium-voltage cable [89].***

With particular focus on the moisture, two different sites have been considered, because providing some significant remark.

In the *SITE A* the measured values are broadly comparable along the entire measurement period and exhibit a fairly stable trend (Figure 10). However, as is clearly visible, a pronounced peak in surface relative humidity occurs, starting in the second half of April and reaching a maximum in the first days of May. This feature is not attributable to a sensor or data-logger malfunction during data acquisition; rather, it corresponds to a period of intense rainfall, with ambient humidity levels reaching peaks of up to 95%. Subsurface moisture is naturally affected as well, but with a delay and with a markedly smaller amplitude.

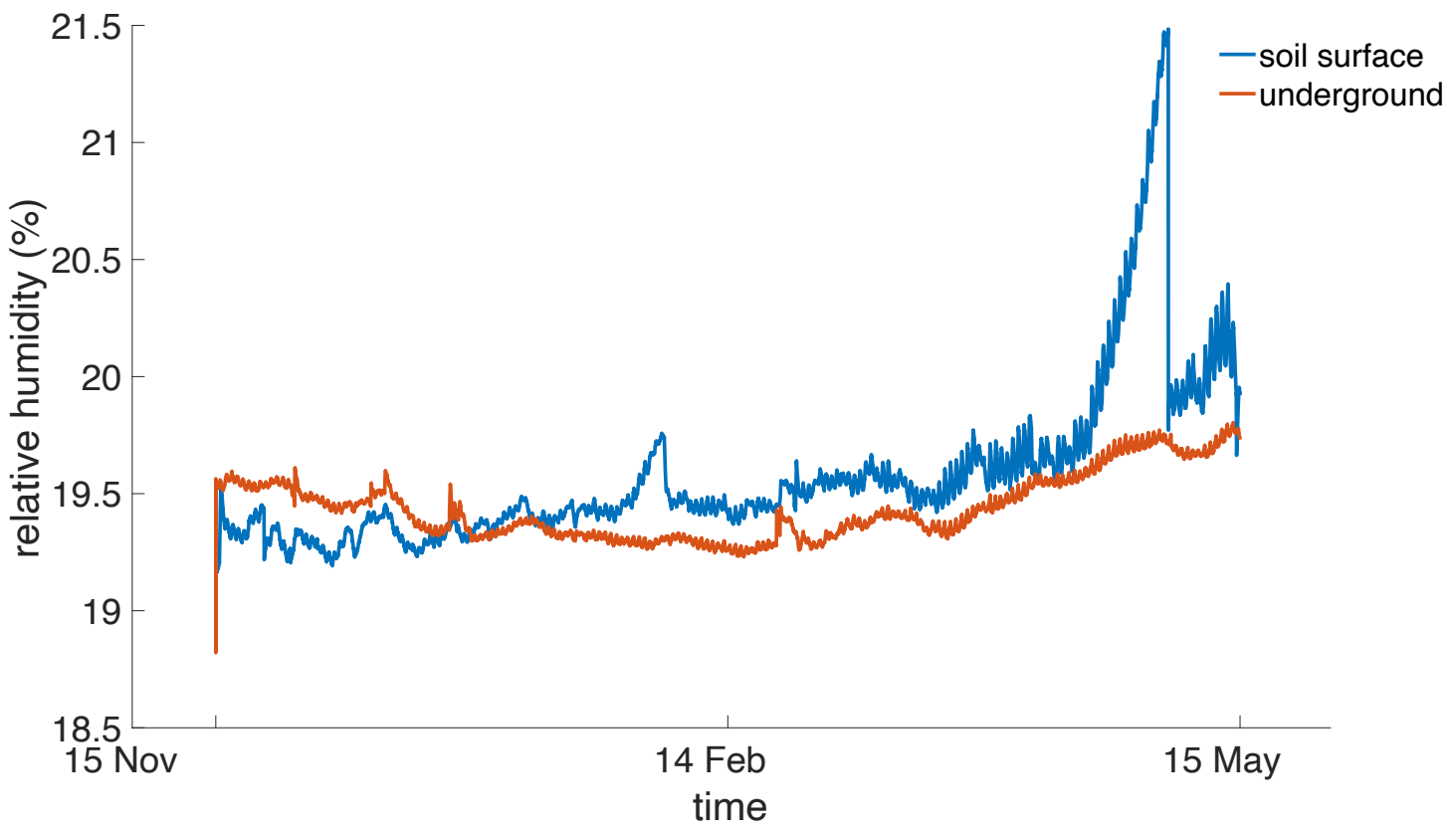

***Figure 10 Measured relative humidity [%].***

The difference in relative humidity is fairly small, on the order of about 0.5%, as it is mainly driven by ambient conditions (Figure 11). Another parameter that plays a role, however, is the soil type in which the cable is installed. The soil considered here, and hence the adopted soil-water retention curve, corresponds to a sandy soil, for which moisture variations are limited. If, instead, a clayey soil had been considered, where moisture content changes more markedly with variations in pore-water pressure, substantially larger discrepancies would be expected.

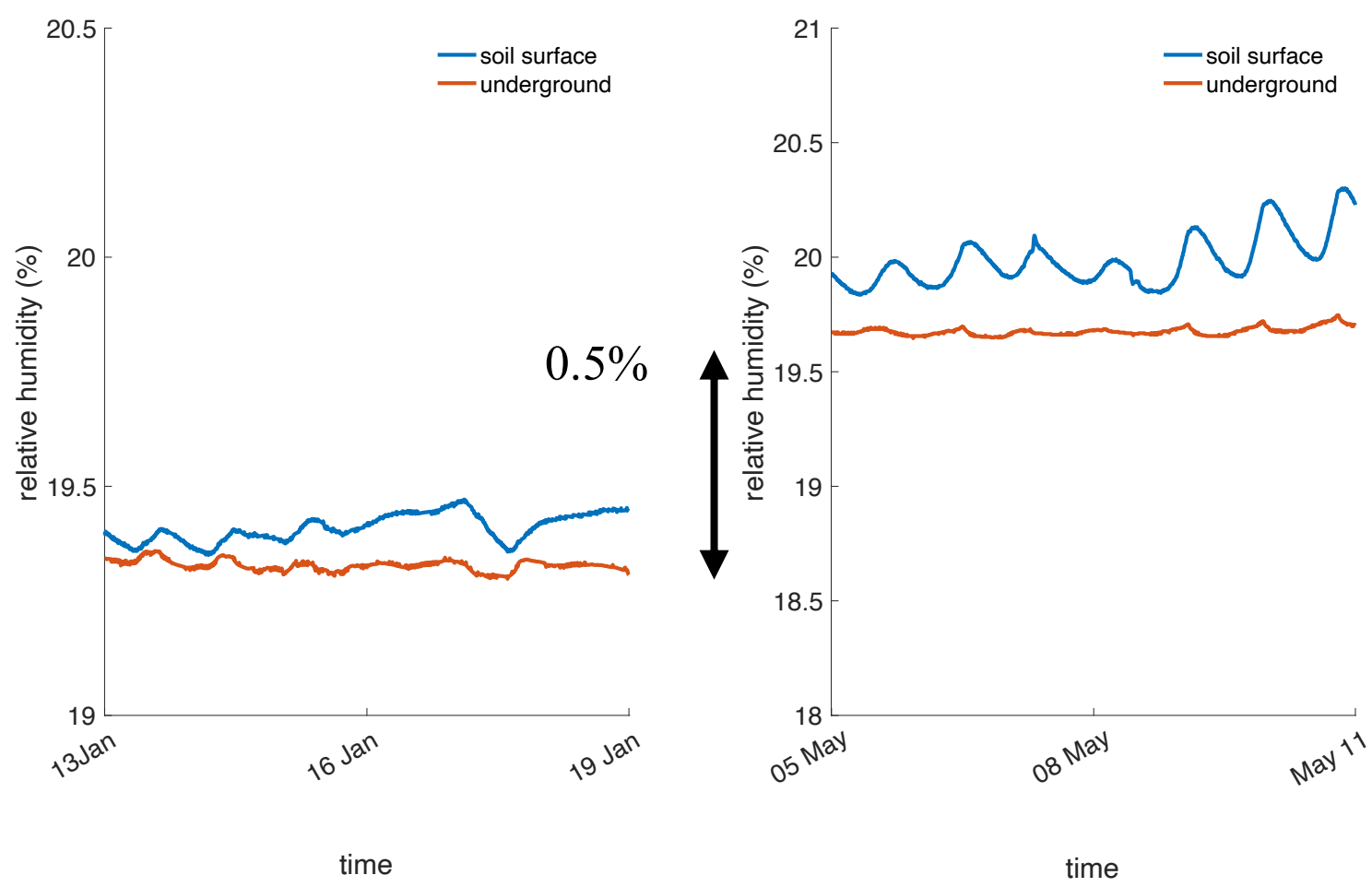

***Figure 11 Comparative trend of relative humidity on the surface of the analysed soil and at depth across different periods of the year.***

A different site, namely *SITE B*, shows a slightly different condition: as shown in Figure 12, the trends of relative humidity at the surface and in the subsurface are again more or less steady, except for irregularities that consistently occur prior to line failures. The most evident feature, however, is the marked alternation between the two humidity signals in terms of which one attains the higher values; this behaviour is clearly influenced by the occurrence (or absence) of precipitation events.

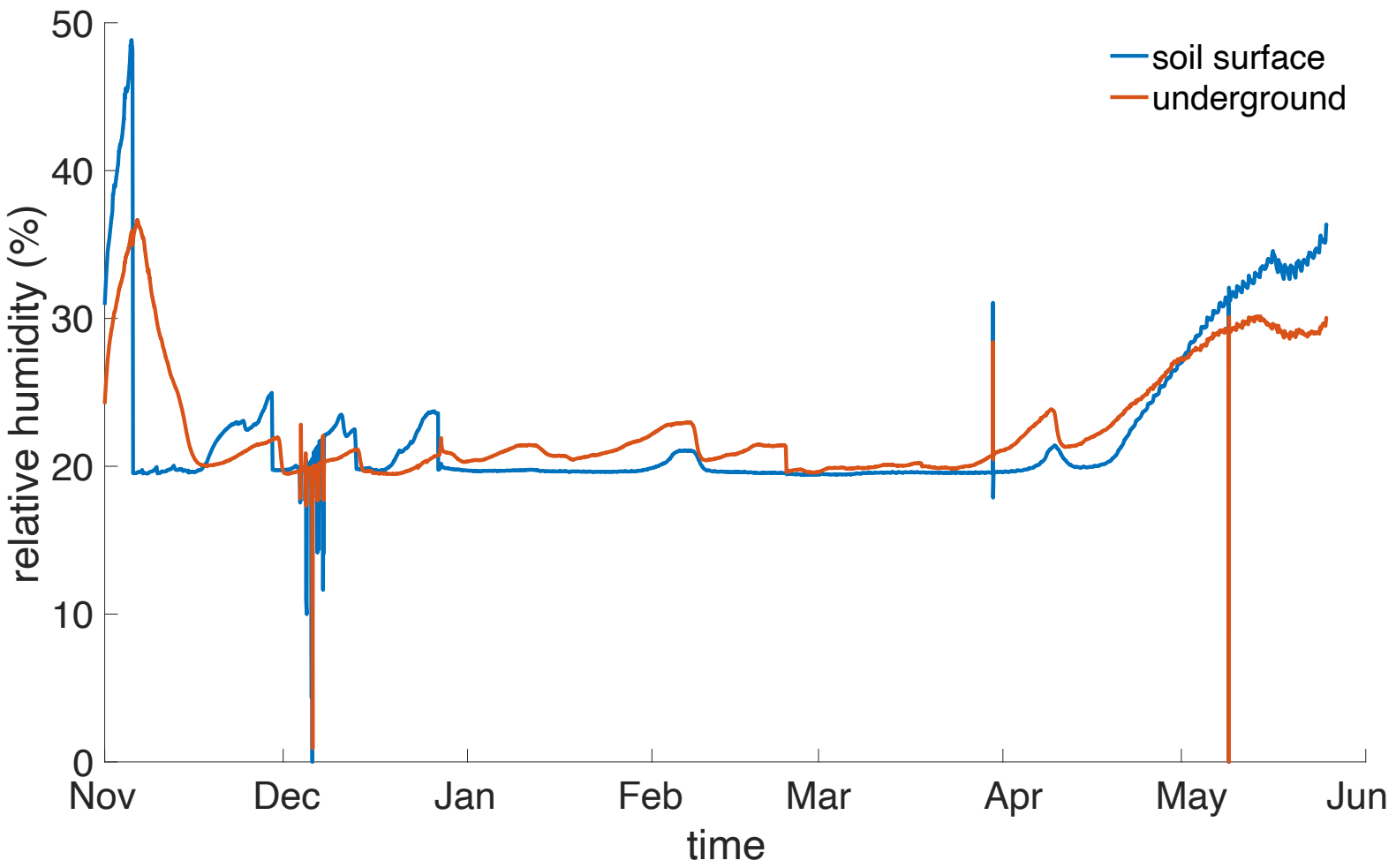

***Figure 12 Comparative trend of relative humidity on the surface of the analysed soil and at depth over a six-month monitoring period.***

Between autumn and spring, relative humidity variations are instead much more pronounced, not only at the surface but also in the subsurface (on the order of about 5%). However, rainfall has a strong influence on the soil and, consequently, on the values recorded by the sensor (Figure 13).

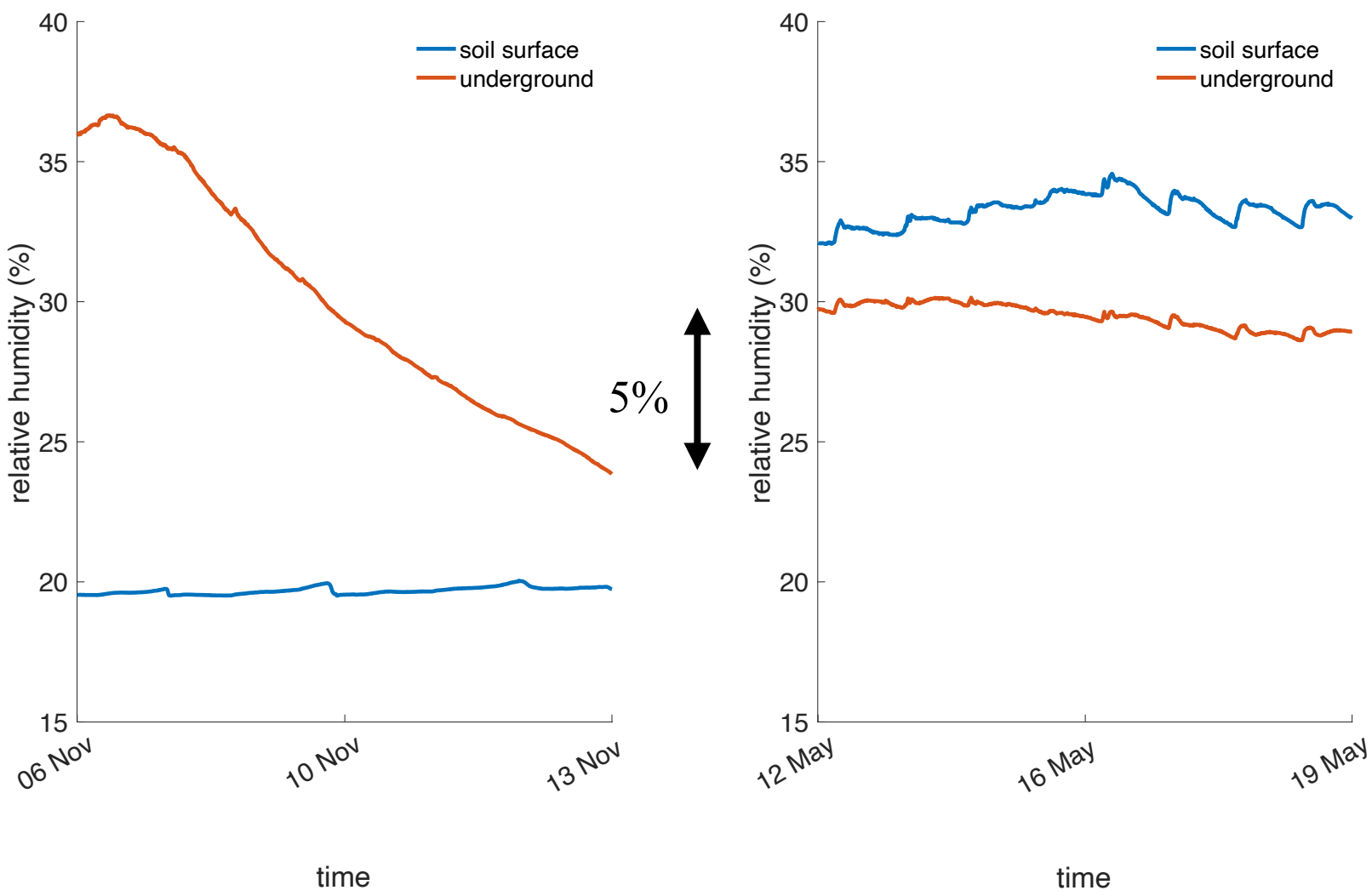

***Figure 13 Comparative trend of relative humidity on the surface of the analysed soil and at depth, focusing on the transition between late winter and early summer.***

### 5.2 *Information on Real World Joint Defects*

From the analysis of several dozen joints taken from MV underground cables (see [7] [20] [22] for specific details), multiple degradation modes were identified, including:

- deterioration of the metallic screen and of the insulation;
- in some cases, workmanship/installation defects;
- presence of voids at the interfaces between dissimilar materials.

Notably, almost all failures involve joints connecting cables with *different insulation technologies*, namely extruded-insulation cables and PILC (transition joints). Examination of the collected joints indicates that most of them experienced an ultimate failure due to degradation of the oil-impregnated paper insulation. Screen degradation was observed in most of the remaining joints, nearly half of which had also experienced an ultimate failure. Such joint-manufacturing/installation imperfections (sometimes unavoidable due to the technological complexity) do not necessarily lead to early failure; rather, they typically require several thermal cycles and electrical stresses before evolving into a definitive breakdown discharge.

In urban areas, trench excavation and cable laying can be carried out in two separate stages. In particular, using plastic ducts enables the excavation of short trench sections that can be rapidly backfilled, thereby minimising disruption to city traffic. Once several sections have been completed, cable installation can then be performed (typically over lengths of a few hundred metres) without affecting surface traffic. However, duct installation is associated with a reduced current-carrying capacity (by up to approximately 15%, all else being equal) compared with direct burial, due to the additional thermal barrier represented by the warm air trapped inside the plastic duct.

### 5.3 *Effect of the Moisture on Terminals in MV/LV Substation*

The effects on cables and joints represent only part of the issues that moisture can cause within the system. During the preparation of this paper, interviews conducted with some DSOs[2] revealed that moisture-related problems have also been encountered in MV substations.

First of all, they noted differences among different terminal providers: this suggests that the surface of the insulators and terminations presents some micro-defects, whose presence may lead to the electrical treeing. Hence, as preventive action, a cleaning procedure has been introduced. It is based on two-step process: during the first step, terminals and insulators are cleaned by using a special paste, in order to eliminate any residuals from the micro-defects. After doing that, the second step implies the filling of the micro-defects through a silicon granular material. This enables to increase the insulation resistance by three orders of magnitude (from MΩ to GΩ). However, it is important to highlight that the correct execution of the first step is fundamental: in fact, if the cleaning is not done properly, the use of the granular material on non-cleaned surfaces worsens the situation (i.e., insulation resistance may fall to tens of kΩ).

[2] The information has been provided by agreeing to keep the confidentiality about the company name.

The moisture on the substation components appears as water drops on the surface of the substation compartments. It has been noted that these drops appear with level of relative humidity around 60%, but with certain temperature conditions. The exact combination of temperature and relative humidity has not been totally addressed yet. It is worth noting that since these phenomena arise from the combined influence of temperature and humidity, secondary substations equipped with an MV/LV transformer have been observed to exhibit a lower incidence of discharge events (the presence of the transformer losses enable a reduction of the humidity level within the substation). For this reason, in particular cases, where the problem was severe, some DSOs have installed heaters/dehumidifiers in order to reduce the relative humidity and hence the probability of fault. These preventive measures have been actually confirmed by enclosing manufacturers, as described in the customer indications shown in [54].

Regarding the type of faults, those caused by moisture in the substation compartment are temporary: in fact, the discharge increases the temperature of the insulator, thereby eliminating the moisture. However, the fact that these faults are temporary does not mean that they are without consequences. In fact, firstly, the fault occurrence leads to an interruption of the supply (decreasing the reliability indexes of the DSO). Secondly, the insulator, after the discharge, usually presents some damages that reduce its insulation level, requiring its replacement.

From what presented above, it is evident how integrated strategies combining material selection, monitoring, diagnostics, and maintenance are crucial to mitigate moisture-related failures in real distribution networks.

## 6 *Conclusions*

The literature review and field experiences discussed in this work show that moisture can affect cables, joints, accessories, and substation equipment through multiple degradation mechanisms. These include corrosion processes, dielectric property modifications, electric-field distortion, partial discharge activity, water-tree formation and propagation, and ultimately insulation breakdown. Although these processes often evolve slowly over many years, their cumulative effect may significantly reduce asset lifetime, increase failure rates, and compromise network reliability and resilience.

The study also confirms that the response to moisture strongly depends on material characteristics. XLPE, EPR, EPDM, and emerging nanocomposite-based insulation systems exhibit different moisture uptake mechanisms and degradation behaviors. Consequently, material selection, moisture barriers, improved joint designs, and advanced formulations specifically engineered to retard water-treeing and moisture diffusion represent effective strategies for enhancing the long-term reliability of MV cable systems.

From a diagnostic perspective, no single technique can provide a complete assessment of moisture-related ageing. Instead, reliable condition evaluation requires a multi-layered diagnostic approach combining dielectric response measurements, partial discharge monitoring, reflectometry techniques, thermal investigations, and laboratory characterization methods. The complementary use of these techniques enables both the detection of external moisture ingress and the identification of internal degradation processes before catastrophic failures occur.

The practical experiences presented in the paper further demonstrate the importance of continuous environmental monitoring and proper installation and maintenance practices. Soil moisture measurements, thermal monitoring of underground assets, improved sealing procedures, surface treatments, and humidity-control measures in MV/LV substations all represent effective tools for reducing moisture-related failures and supporting condition-based maintenance strategies.

Overall, the main contribution of this work is the integrated analysis of moisture effects from the material scale up to the network-operational level. The findings highlight that effective moisture management should not be regarded as a purely materials issue or a maintenance issue alone, but rather as a system-wide reliability challenge requiring coordinated actions in design, installation, monitoring, diagnostics, and asset management. As power distribution networks continue to expand and evolve to support the energy transition, improving the understanding, detection, and mitigation of moisture-related degradation will be essential for extending asset lifetime, reducing outages, and ensuring the long-term resilience of underground MV infrastructure.

Future research should focus on the development of digital monitoring solutions, moisture-aware asset health models, advanced moisture-resistant insulation materials, and data-driven diagnostic methods capable of combining environmental, thermal, and electrical information into unified condition assessment frameworks. Such developments may enable a further transition from preventive maintenance practices toward predictive and risk-based asset management of distribution networks.

**Acknowledgements**

This study was partially developed, as affine topic, within the project EXTRASTRONG (resilience evaluation by EXperimental and TheoRetical ApproacheS in electrical distributiON systems with underGround cables) – funded by European Union – Next Generation EU within the PRIN 2022 program (D.D.104 – 02/02/2022 – Ministero dell'Università e della Ricerca).
This manuscript reflects only the authors' views and opinions, and the Ministry cannot be considered responsible for them.

**Declarations**

*Conflict of interest* The authors declare no competing interests.
*Ethics approval and consent to participate* Not applicable.
*Consent for publication* Not applicable.